\documentclass[aps,prd,preprint,showpacs]{revtex4}
\usepackage{graphicx}

\begin{document}

\title{Post-Newtonian Models of Differentially Rotating Neutron Stars}
\label{ch:pn}

\author{\firstname{Yuk Tung} \surname{Liu}}
\affiliation{Theoretical Astrophysics, California Institute of Technology,
Pasadena, California 91125}
\altaffiliation[Address after August 20, 2002: ]{Department of Physics, University of 
Illinois at Urbana-Champaign, Urbana, IL~61801}
\email{ytliu@tapir.caltech.edu}

\date{\today}

\begin{abstract}
A self-consistent field method is developed, which can be used to construct 
models of 
differentially rotating stars to first post-Newtonian order. The rotation 
law is specified by the specific angular momentum distribution 
$j(m_{\varpi})$, where $m_{\varpi}$ is the baryonic mass fraction 
inside the surface of constant specific angular momentum. The method 
is then used to compute models of the nascent neutron stars resulting from 
the accretion induced collapse of white dwarfs. The result shows that 
the ratios of kinetic energy to gravitational binding energy, 
$\beta$, of the relativistic models are slightly smaller than the 
corresponding values of the Newtonian models. 
\end{abstract}


\maketitle

\section{Introduction}

In a recent paper~\cite{liu02} (hereafter Paper~II), we demonstrate that 
the accretion induced 
collapse (AIC) of a rapidly rotating white dwarf can result in a rapidly 
rotating neutron star that is dynamically unstable to the bar-mode 
instability, which is the instability resulting from non-axisymmetric perturbations 
with angular dependence $e^{\pm 2i\varphi}$. Here $\varphi$ is the azimuthal 
angle. This instability could emit a substantial amount 
of gravitational radiation that could be detectable by gravitational 
wave interferometers, such as LIGO, VIRGO, GEO and TAMA. 

However, for this instability to occur, the neutron 
star must have a $\beta = T/|W|$ greater than a critical 
value $\beta_d\approx 0.25$ (Paper~II). Here $T$ is 
the rotational kinetic energy and $|W|$ is the gravitational 
binding energy. Only the AIC of white dwarfs that are composed of 
oxygen, neon and magnesium (O-Ne-Mg white dwarfs) with 
$\Omega > 0.93 \Omega_m$ can produce neutron stars with such a high 
value of $\beta$. Here $\Omega$ is the angular velocity of the white 
dwarf and $\Omega_m$ is the angular velocity at which mass shedding 
occurs on the equatorial surface. This type of source will not be 
promising for LIGO~II because its event rate is not expected to be very high. 

Neutron stars are compact objects. General relativistic effects have  
a significant influence on both the structure and dynamical stability 
of the stars. Recently, Shibata, Baumgarte, Saijo and Shapiro studied 
the dynamical stability of differentially rotating polytropes in full 
general relativity~\cite{shibata00} and in the post-Newtonian 
approximation~\cite{saijo00}. 
They performed numerical simulations on the differentially rotating 
polytropes with some specified rotation law. They found that as the 
star becomes more compact, the critical value $\beta_d$ slightly 
decreases from the Newtonian value $0.26$ to $0.24$ 
for their chosen rotation 
law. It is not clear, however, whether their result implies that 
relativistic effects would destabilize the stars we are studying, 
for the equilibrium 
structure of the star will also be changed by relativistic effects. The 
value of $\beta$ of a relativistic star will not be the same as 
that of a Newtonian star with the same baryon mass and total angular 
momentum. 

The objective of this paper is twofold. First, we develop a new numerical 
technique to construct the equilibrium structure of a rotating star with 
a specified specific angular momentum distribution to first 
post-Newtonian (1PN) order [i.e., including terms of order $c^{-2}$ higher
than the Newtonian terms in the equations of motion]. Then we use this new 
technique to construct models of neutron stars 
corresponding to the collapse of the white dwarfs we studied in Paper~II 
and Ref.~\cite{liu01} (hereafter Paper~I) and compare them with the 
Newtonian models. 

Equilibrium models of neutron stars in full general relativity have
been built by many authors~\cite{wilson72,bonazzola74,butterworth75,butterworth76,butterworth7679,friedman86,komatsu89}.
The neutron stars studied in the literature are either rigidly rotating or rotating
with an \textit{ad hoc} rotation law. New-born neutron stars resulting from
core collapse of massive stars or accretion induced collapse of massive
white dwarfs are differentially
rotating~\cite{monchmeyer88,janka89,fryer01,liu01,liu02}.
It seems plausible that the rotation laws of these neutron stars could be
approximated by the specific angular momentum distribution $j(m_{\varpi})$
of the pre-collapse stars (see Paper~I and Sec.~\ref{sec:fullGR}).
Here $m_{\varpi}$ is the baryonic mass fraction
inside the surface of constant specific angular momentum.
Equilibrium models of Newtonian stars with a specified $j(m_{\varpi})$
have been constructed by many 
authors~\cite{ostriker68,bodenheimer73,pickett96,new01}.
However, none of these studies, to our knowledge, has been generalized
to include the relativistic effects. 

If a rotating axisymmetric star is described by a barotropic equation 
of state, i.e., the total energy density $\epsilon$ is a function of 
pressure only, then there is a constraint on the rotation law (see 
Section~\ref{sec:fullGR}). This rotational constraint is usually written in 
the form $u^0 u_{\varphi}=F(\Omega)$~\cite{bardeen70,butterworth76}, 
where $F$ is an arbitrary function. 
Here $\Omega$ is the angular velocity of the fluid with respect to 
an inertial observer at infinity; $u^0$ is the time component of the 
four-velocity and $u_{\varphi}=u_{\mu} \varphi^{\mu}$, 
where $\varphi^{\alpha}$ is the axial Killing vector field of the 
spacetime. In the Newtonian limit, this constraint reduces to the 
well-known result that $\Omega$ is constant in the direction parallel 
to the rotation axis. The major obstacle in the construction of 
differentially rotating relativistic stars is that it is not clear what 
function $F$ should be used to produce the desired specific angular 
momentum distribution $j(m_{\varpi})$. In the next section, we will 
reformulate the rotational constraint in a way that can be used to 
impose the rotation law $j(m_{\varpi})$, at least in the 1PN calculations.

The structure of this paper is as follows. In Section~\ref{sec:form}, 
we give a brief review on the full relativistic treatment of rotating 
relativistic stars and then reformulate the rotational constraint imposed 
by the barotropic equation of state. Next, we 
use the standard 1PN metric and show that the rotational 
constraint can be solved analytically. We then derive the equations 
of motion determining the structure of a star to 1PN 
order. In Section~\ref{sec:num}, we generalize the self-consistent 
field method of Smith and Centrella~\cite{smith92} so that it can be used to
compute the structure of a star to 1PN order. In Section~\ref{sec:results}, 
we apply the numerical method to construct neutron star models resulting 
from the collapse of the O-Ne-Mg white dwarfs we studied in Paper~II
and compare them with the corresponding Newtonian models. Finally, 
we summarize our conclusions in Section~\ref{sec:con}.

\section{\label{sec:form}Formalism}

In this section, we first give a brief review on the full relativistic 
treatment of rotating relativistic stars and then reformulate the rotational 
constraint imposed by the barotropic equation of state (EOS) in 
Section~\ref{sec:fullGR}. Then we derive the equations of 
motion determining the equilibrium structure of a rotating star 
to 1PN order in Section~\ref{sec:pn}.
Throughout this chapter, we 
use the convention that Greek indices run from 0 to 4, 0 being the 
time component; whereas Latin indices run from 1 to 3 only. A sum over 
repeated indices is assumed unless stated otherwise. The signature of the 
metric is $(-+++)$. 

\subsection{\label{sec:fullGR}Relativistic hydrodynamics}

We want to construct the nascent neutron stars resulting from the AIC 
of rotating white dwarfs. As in Paper~I, we make the following assumptions 
on the AIC and the collapsed stars.

First, we assume the collapse is axisymmetric. Hence the spacetime, albeit 
dynamical, has an axial Killing vector field $\varphi^{\alpha}$. Second, 
we neglect viscosity and assume a perfect fluid stress-energy tensor 
\begin{equation}
  T^{\mu \nu} = (\epsilon + P) u^{\mu} u^{\nu}+Pg^{\mu \nu} \ ,
\label{Tab}
\end{equation}
where $\epsilon$ is the energy density in the fluid's rest frame; $P$ is 
pressure and $u^{\mu}$ is the fluid's four-velocity, normalized so that 
$u^{\alpha} u_{\alpha}=-1$. 
Third, we assume that the collapsed objects can 
be described by a barotropic EOS, i.e.\ $\epsilon=\epsilon(P)$. 
Fourth, we assume there is no meridional circulation in the equilibrium state 
of the collapsed objects, i.e., the spacetime is \textit{nonconvective} 
or \textit{circular}~\cite{gourgoulhon93}. In other words, the fluid's 
four-velocity can be written as 
\begin{equation}
  u^{\alpha} = u^0 \left(t^{\alpha} + \frac{\Omega}{c} \varphi^{\alpha}\right) \ ,
\label{4-u}
\end{equation}
where $t^{\alpha}$ is the timelike Killing vector field of the spacetime of
the collapsed star, $c$ is the speed of light, $u^0$ is the time component 
of the four-velocity, and $\Omega$ is the rotational angular velocity measured 
by an inertial observer at infinity.
Finally, we assume that no material is ejected from the star during and 
after the collapse. Hence the total baryon rest mass $M_0$ and the total 
angular momentum $J$ are conserved.

Let $n$ denote the baryon number density in the fluid's rest frame. It 
follows from the baryon number conservation law $\nabla_{\nu} (n u^{\nu})=0$ 
and conservation of energy-momentum $\nabla_{\nu} T^{\mu \nu}=0$ that 
(see, e.g., Chapter~22 of Ref.~\cite{MTW})
\begin{equation}
  \frac{d\epsilon}{dn} = \frac{\epsilon+P}{n} \ .
\end{equation}
Given a barotropic EOS, the above equation can be integrated, giving 
\begin{equation}
  n(\epsilon) = n(\epsilon_0)\, \exp \left[ \int_{\epsilon_0}^{\epsilon} 
\frac{d\epsilon'}{\epsilon'+P(\epsilon')} \right] \ .
\end{equation}
We define the baryonic rest mass density $\rho = n \bar{m}_B$.
Here $\bar{m}_B$ is the average baryon mass, defined so that 
\begin{equation}
  \lim_{\epsilon \rightarrow 0} \frac{\epsilon}{\rho c^2}=1 \ .
\end{equation}
It follows~\cite{bardeen70} from the conservation of 
baryon number $\nabla_{\nu} (n u^{\nu})=0$, conservation of energy-momentum 
$\nabla_{\nu} T^{\mu \nu}=0$, and the existence of an axial Killing 
vector $\varphi^{\alpha}$ that 
\begin{equation}
  u^{\beta} \nabla_{\beta} j = \frac{dj}{d\tau}=0 \ ,
\label{jcon}
\end{equation}
where $\tau$ is the proper time along the fluid particle's worldline and the 
specific angular momentum 
\begin{equation}
  j = \frac{\epsilon+P}{\rho c} u_{\varphi} \ .
\label{jdef}
\end{equation}
Here $u_{\varphi}=u_{\alpha}\varphi^{\alpha}$. 
In the Newtonian limit, $j=\Omega \varpi^2$, which is the Newtonian expression 
of the specific angular momentum along the rotation axis. Here $\varpi$ is 
the distance from the rotation axis. Following the arguments in Paper~I, we 
conclude that the final neutron star should have the same baryon mass 
$M_0$, total angular momentum $J$, and specific angular
momentum distribution $j(m_{\varpi})$ as the pre-collapse white dwarf.

In the stationary and axisymmetric spacetime of a relativistic star, 
the Euler equation takes the form~\cite{bardeen70} 
\begin{equation}
  \frac{\nabla_{\alpha} P}{\epsilon+P}=-u^{\beta}\nabla_{\beta}u_{\alpha} 
= \nabla_{\alpha} (\ln u^0)-u^0 u_{\varphi}\frac{\nabla_{\alpha} \Omega}{c} \ .
\label{Eulereq}
\end{equation}
Since the EOS is barotropic, the left side of Eq.~(\ref{Eulereq}) is 
a total differential. This imposes a constraint on the rotation law: the 
integrability condition for Eq.~(\ref{Eulereq}) is that the rotation 
law must have the form 
$u^0u_{\varphi}=F(\Omega)$~\cite{bardeen70,butterworth76}, where $F$ 
is an arbitrary 
function. In the Newtonian limit, this rotational constraint means that $\Omega$ is 
constant in the direction parallel to the rotation axis. The constraint 
written in this form is not convenient for our purpose, as our rotation 
laws are specified by the function $j(m_{\varpi})$. Hence, we formulate 
the constraint in another way: the integrability condition is that 
$u^0 u_{\varphi}\nabla_{\alpha} \Omega$ is an exact differential. In 
the language of differential forms, we require that $u^0 u_{\varphi}\tilde{d} 
\Omega$ be an exact form. This implies that its exterior derivative 
vanishes: 
\begin{equation}
  \tilde{d}(u^0 u_{\varphi} \tilde{d}\Omega)=0 \ ,
\label{rot-con}
\end{equation}
where $\tilde{d}$ denotes the exterior derivative.

\subsection{\label{sec:pn}Post-Newtonian approximation}

Following Chandrasekhar~\cite{chandra65}, we split the energy density into two terms:
\begin{equation}
  \epsilon = \rho c^2 \left( 1+\frac{\Pi}{c^2}\right) \ .
\label{defPI}
\end{equation}
We adopt the 1PN metric (in Cartesian coordinates) developed 
by Chandrasekhar, and Blanchet, Damour and 
Sch\"afer~\cite{chandra65,blanchet90,cutler91}: 
\begin{eqnarray}
  g_{00} &=& -\left(1+\frac{2U}{c^2}+\frac{2U^2}{c^4}\right) + O(c^{-6}) 
\ , \label{g00} \\
  g_{0i} &=& \frac{A_i}{c^3}+O(c^{-5}) \ , \label{g0i} \\
  g_{ij} &=& \left( 1-\frac{2U}{c^2}\right) \delta_{ij}+O(c^{-4}) \ . 
\label{gij}
\end{eqnarray}
The metric components satisfy the gauge condition 
\begin{eqnarray}
  \sum_{i=1}^3 \partial_i g_{i0}-\frac{1}{2c}\partial_t \left(\sum_{i=1}^3 g_{ii}
\right) &=& O(c^{-5}) \ , \\ 
\sum_{i=1}^3 \partial_i g_{ij}+\frac{1}{2}\partial_j \left( g_{00}-\sum_{i=1}^3 
g_{ii}\right) &=& O(c^{-4}) \ .
\end{eqnarray}
In this metric, the components of the four-velocity are 
\begin{eqnarray}
  u^0 &=& 1+\frac{1}{c^2} \left( \frac{v^2}{2}-U\right)+
\frac{1}{2c^4}\left(U^2-5Uv^2+\frac{3}{4}v^4+2v^i A_i\right) 
  +O(c^{-6}) \ , \label{u0} \\
  u^i &=& \frac{v^i}{c}\left[ 1+\frac{1}{c^2}\left(\frac{v^2}{2}-U\right)\right] 
+ O(c^{-5}) \ , \label{ui}
\end{eqnarray}
where $v^i=c u^i/u^0$ and $v^2=\delta_{ij} v^i v^j$.
The potentials $U$ and $A_i$ satisfy the elliptic equations
\begin{eqnarray}
  D_j D^j U + \frac{8\pi G\rho}{c^2}U &=& 4\pi G \rho \left[ 
1+\frac{1}{c^2}\left( \Pi+2v^2+\frac{3P}{\rho}\right)\right] \ , 
\label{Ueq} \\
  D_j D^j A_i &=& 16\pi G \rho \delta_{ik} v^k \ , \label{Aeq}
\end{eqnarray}
where $D_j$ denotes the covariant derivative compatible with the three 
dimensional flat-space metric, and $G$ is the gravitational constant. 
We introduce cylindrical coordinates $(\varpi,\varphi,z)$ with 
$\partial/\partial \varphi$ being the axial Killing vector, and 
$\varpi = \sqrt{x^2+y^2}$. In this coordinate system, the velocity 
vector potential $A_i$ has only one component: $A_{\varphi}=xA_y-yA_x$. 
Let $Q= A_{\varphi}/\varpi$. Then $Q$ satisfies the equation 
\begin{equation}
  D^j D_j Q - \frac{Q}{\varpi^2} = 16\pi G \rho \Omega \varpi \ .
\label{Qeq}
\end{equation}

To 1PN order, we have 
\begin{equation}
  c u^0 u_{\varphi} = \varpi^2 \Omega \left( 1+\frac{K}{c^2}\right) \ , 
\end{equation}
where 
\begin{equation}
  K = v^2-4U+\frac{Q}{v} \ .
\label{Kdef}
\end{equation}
Since $u^0 u_{\varphi}\nabla_{\alpha} \Omega$ is an exact differential, 
we can write
\begin{equation}
  \nabla_{\alpha} f = \varpi^2 \Omega \left(1+\frac{K}{c^2}\right) 
\nabla_{\alpha}\Omega \ ,
\label{fdef}
\end{equation}
where $f$ is a scalar function. 
The rotational constraint~(\ref{rot-con}) gives only one nontrivial 
equation for $\Omega$ in a stationary and axisymmetric spacetime. 
To 1PN order, Eq.~(\ref{rot-con}) can be solved analytically, giving 
\begin{equation}
  \Omega(\varpi,z)=\Omega_0(\varpi)+\frac{\varpi \partial_{\varpi}\Omega_0}{2c^2}
[K(\varpi,z)-K_0(\varpi)] \ ,
\label{omega}
\end{equation}
where $\Omega_0(\varpi)\equiv \Omega(\varpi,0)$ and $K_0(\varpi)\equiv K(\varpi,0)$. 
Eq.~(\ref{fdef}) can then be integrated and we obtain, up to an arbitrary 
additive constant,
\begin{eqnarray}
  f(\varpi,z) &=& \frac{\varpi^2}{2}\Omega_0^2(\varpi)-\int_0^{\varpi}
d\varpi' \, \varpi' \Omega_0^2(\varpi') \cr
 & & +\frac{1}{c^2}\left[ 
\varpi^2 \Omega_0(\varpi)\, \Omega_1(\varpi,z)+\frac{\varpi^2}{2}\Omega_0^2(\varpi) 
K_0(\varpi)-I_1(\varpi)\right] \ , \ \ \ \ \ \ \ \ 
\end{eqnarray}
where 
\begin{eqnarray}
  \Omega_1(\varpi,z) &=& \frac{\varpi \partial_{\varpi}\Omega_0}{2}
[K(\varpi,z)-K_0(\varpi)] \ , \label{omega1} \\
  I_1(\varpi) &=& \int_0^{\varpi} d\varpi' \, \Omega_0^2(\varpi') 
\left[ \varpi' K_0(\varpi')+\frac{\varpi'^2}{2}\partial_{\varpi} K_0(\varpi') 
\right] \ .
\end{eqnarray}

It is convenient to define an auxiliary function 
\begin{equation}
  h=c^2 \int_0^P \frac{dP'}{\epsilon(P')+P'} \ .
\label{hdef}
\end{equation}
This quantity is defined only inside the star. The boundary of the star 
is given by the surface $h=0$. In the Newtonian limit, $h$ reduces to 
the specific enthalpy. The Euler equation~(\ref{Eulereq}), to
1PN order, can be written in integral form:
\begin{eqnarray}
  h(\varpi,z) &=& \int_0^{\varpi} d\varpi' \, \varpi' \Omega_0^2(\varpi') 
- U +C \cr
  & & +\frac{1}{c^2}\left( \frac{\varpi^4}{4}\Omega_0^4 
-2\varpi^2 \Omega_0^2U+Qv-\frac{\varpi^2\Omega_0^2K_0}{2} 
+I_1 \right) \ , \ \ \ \ \ 
\label{Euler1pn}
\end{eqnarray}
where $C$ is a constant and all quantities outside the integral are evaluated 
at $(\varpi,z)$.

The structure of the star is determined once a rotation law is given. 
The rotation law is specified by the specific angular momentum 
distribution function 
$j(m_{\varpi})$, which is determined by the pre-collapse white dwarf 
(see Paper~I). Straightforward calculations using 
Eqs.~(\ref{jdef}), (\ref{defPI}), (\ref{g00}), (\ref{g0i}), (\ref{gij}), 
(\ref{u0}), (\ref{ui}), (\ref{omega}) and~(\ref{omega1}) give 
\begin{equation}
  j = \varpi^2 \Omega_0 \left[ 1+\frac{1}{c^2}\left(\frac{v^2}{2}-3U
+\Pi+\frac{P}{\rho}+\frac{\Omega_1}{\Omega_0}+\frac{Q}{v}\right) 
\right] \ .
\label{j1pn}
\end{equation}
To compute $m_{\varpi}$, the baryonic mass fraction inside the surface 
of constant $j$, we first need to determine the surfaces on which $j$ 
is constant. 
In the Newtonian case, the surfaces of constant $j$ are cylinders. This is not 
true in general in the relativistic case (at least not in the coordinate 
system we are using). 
Let $[\varpi+\eta(\varpi,z)/c^2,z]$ denote the surface of constant $j$ that 
intersects the equatorial plane at cylindrical coordinate radius $\varpi$. 
Hence we have 
\begin{eqnarray}
  j[\varpi+\eta(\varpi,z)/c^2,z] &=& j(\varpi,0) \ , \label{jconstant} \\ 
  \eta(\varpi,0)=0 \ .
\end{eqnarray}
Expanding the left side of Eq.~(\ref{jconstant}) to $O(c^{-2})$, we 
obtain 
\begin{equation}
  \eta(\varpi,z)=-c^2 \frac{j(\varpi,z)-j_0(\varpi)}{\partial_{\varpi} 
j_0(\varpi)} \ ,
\end{equation}
where $j_0(\varpi)\equiv j(\varpi,0)$. Using Eq.~(\ref{j1pn}), we 
obtain 
\begin{equation}
  \eta = -\frac{\varpi \Omega_0 \pounds (\Pi-3U+P/\rho)+\varpi \Omega_1 
+\pounds (Q)}{2\Omega_0+\varpi \partial_{\varpi}\Omega_0} \ ,
\label{eta}
\end{equation}
where $\pounds (q) \equiv q(\varpi,z)-q(\varpi,0)$. The baryon mass 
$M_{\varpi}$ inside the volume $V_{\varpi}$ bounded by the surface of 
constant $j$ is given by
\begin{eqnarray}
  M_{\varpi} &=& \int_{V_{\varpi}} \rho u^{\mu} n_{\mu}\,
dV \\
 &=& 2\pi\int_{-\infty}^{\infty} dz'\, \left[ \varpi 
\rho^*(\varpi,z') \frac{\eta(\varpi,z')}{c^2}+\int_0^{\varpi}d\varpi' \, 
\varpi' \rho^*(\varpi',z') \right] \ , \ \ \ \ \ \ 
\label{Mvarpi}
\end{eqnarray}
where $n^{\mu}$ 
is the unit vector orthogonal to the surface of constant $t$; $dV$ is the 
proper volume element in the constant $t$ hypersurface, and 
\begin{equation}
  \rho^*=\rho \left[ 1+\frac{1}{c^2}\left( \frac{v^2}{2}-3U \right)\right] \ .
\label{rhostar}
\end{equation}
The baryonic mass fraction is then given by 
\begin{equation}
  m_{\varpi}=\frac{M_{\varpi}}{M_0} \ ,
\label{mfraction}
\end{equation} 
where $M_0$ is simply the value of $M_{\varpi}$ at $\varpi=R_e$, the 
equatorial radius of the star.
It is convenient to define the normalized specific angular momentum 
\begin{equation}
  j_n = \frac{M_0}{J} j \ .
\label{jndef}
\end{equation} 
Straightforward calculations from Eq.~(\ref{j1pn}) give 
\begin{equation}
  \Omega_0^2=\frac{\lambda^2 j_n^2}{\varpi^4}-\frac{1}{c^2} 
\left[ \Omega_0^2 \left( \varpi^2\Omega_0^2-6U_0+2\Pi_0+
\frac{2P_0}{\rho_0}\right)+\frac{2\Omega_0Q_0}{\varpi}\right] \ ,
\label{jomega0}
\end{equation}
where $\lambda=J/M_0$ and the subscript ``0'' in the above equation 
means that the quantity is evaluated at $(\varpi,0)$.
The integrated Euler equation~(\ref{Euler1pn}) becomes 
\begin{equation}
  h=\lambda^2 \psi-U+C+\frac{1}{c^2}\left(\frac{v^4}{4}-2Uv^2+Qv-\frac{1}{2}v^2K_0 
+I_2\right) \ .
\label{Eulereq2}
\end{equation}
Here 
\begin{eqnarray}
  \psi(\varpi)&=&\int_0^{\varpi} \frac{j_n^2(m_{\varpi'})}{\varpi'^3}\, 
d\varpi' \ , \label{psi} \\
  I_2(\varpi)&=&\int_0^{\varpi}d\varpi' \, \left[ \frac{1}{2}v_0^2 
\partial_\varpi K_0+2\varpi'\Omega_0^2 \left(U_0-\Pi_0-\frac{P_0}{\rho_0}
\right)-Q_0 \Omega_0\right] \ , \ \ \ \ \ \ \ \label{I2}
\end{eqnarray}
where all the quantities in the integrands are evaluated at $\varpi=\varpi'$.

The rotational kinetic energy $T$ and gravitation potential energy $W$
of a relativistic star are given by (see, e.g.,~\cite{komatsu89})
\begin{eqnarray}
  T &=&\frac{1}{2}\int \Omega\, dJ = \frac{1}{2c}\int \Omega T_{\mu \nu}
n^{\mu} \varphi^{\nu}\, dV \ , \\
  W &=& -[(M_p-M)c^2+T] \ ,
\end{eqnarray}
where the proper mass $M_p$ and gravitational mass $M$ are
\begin{eqnarray}
  M_p &=& \frac{1}{c^2}\int \epsilon u^{\mu} n_{\mu}\, dV \\
  M &=& -\frac{2}{c^2}\int \left( T_{\alpha \beta}-\frac{1}{2}
T^{\sigma}_{\sigma}g_{\alpha \beta}\right) t^{\alpha} n^{\beta}\, dV \ .
\end{eqnarray}
Both $T$ and $W$ are independent of gauge in a spacetime that is 
stationary, axisymmetric and nonconvective.
The expressions for $T$ and $W$ to 1PN order are
\begin{eqnarray}
  T &=& \int \frac{1}{2}\rho v^2 \left[ 1+\frac{1}{c^2} \left(v^2-6U+\Pi
+\frac{P}{\rho}+\frac{Q}{v}\right)\right] \, d^3x \ , \label{eq:pnT} \\
  W &=& \int (\rho v^2+\rho U +3P)\, d^3x -\frac{1}{c^2}\int \rho
\left( \frac{5}{2}U^2+\frac{11}{2}Uv^2-\frac{9}{8}v^4 \right. \cr
  & & \left. -\frac{Qv}{2}-\Pi v^2-\Pi U
-\frac{3Pv^2}{2\rho}+\frac{6PU}{\rho}\right)\, d^3x \ , \label{eq:pnW}
\end{eqnarray}
where $d^3x \equiv \varpi\, d\varpi\, dz \, d\varphi$. 

Given the total baryon mass $M_0$, total angular momentum $J$, normalized 
specific angular 
momentum distribution $j_n(m_{\varpi})$, and EOS, we have to solve 
Eqs.~(\ref{Ueq}), (\ref{Qeq}), (\ref{Eulereq2}), (\ref{j1pn}), 
and~(\ref{mfraction}) consistently to determine the structure of the 
differentially rotating star. We shall discuss how these equations 
can be solved numerically in the next section.

\section{\label{sec:num}Numerical method}

In this section, we develop a self-consistent field technique to calculate  
the structure of a relativistic star with the rotation law specified by 
the normalized specific angular momentum distribution $j_n(m_{\varpi})$. 
Our method is a 
generalization of the one used by Smith and Centrella~\cite{smith92}, 
which is a modified version of Hachisu's self-consistent field 
method~\cite{hachisu86}.

The self-consistent field method is an iteration procedure. Suppose 
in a certain iteration step, we have $h(\varpi,z)$ and  
$\Omega_0(\varpi)$ in a cylindrical grid, we first evaluate the 
quantities $\rho$, $P$ and $\Pi$ from the EOS. Then we compute the potentials 
$U$ and $Q$ by solving the elliptic equations~(\ref{Ueq}) and~(\ref{Qeq}). 
Since the velocity potential $Q$ always appears in the 1PN 
terms of the equations of motion, we can replace $\Omega$ on the right side 
of Eq.~(\ref{Qeq}) by $\Omega_0$. The angular velocity $\Omega$, as well 
as $v=\varpi \Omega$, outside the equatorial plane are determined by 
Eqs.~(\ref{Kdef}) and~(\ref{omega}). Next, we compute the baryonic 
mass fraction $m_{\varpi}$ using Eqs.~(\ref{eta}), (\ref{Mvarpi}),
(\ref{rhostar}) and~(\ref{mfraction}). The function  
$\psi$ is then calculated by Eq.~(\ref{psi}). 
During each iteration, we fix two parameters, which we choose to be  
the central energy density $\epsilon_c$ [or equivalently, $h_c=h(0,0)$] 
and the equatorial radius $R_e$. 
The constants $C$ and $\lambda^2$ in Eq.~(\ref{Eulereq2}) are then 
given by 
\begin{eqnarray}
  C &=& h_c+U_c \ , \\
  \lambda^2 &=& \frac{1}{\psi}\left[ U-C-\frac{1}{c^2}
\left(\frac{v^4}{4}-2v^2U+Qv-\frac{v^2K_0}{2}+I_2\right)\right] \ ,
\end{eqnarray}
where $U_c=U(0,0)$ and all the quantities in the second 
equation are evaluated at the equatorial surface of the star. Finally, 
we update $h$ by Eq.~(\ref{Eulereq2}) and update $\Omega_0$ by 
solving the algebraic equation~(\ref{jomega0}). We repeat the procedure until $h$ 
and $\Omega_0$ converge to the desired accuracy. 

When the star becomes flattened, the iteration scheme described above 
does not converge. This is fixed by generalizing the modified scheme suggested 
in Ref.~\cite{pickett96}: the variables $h$ and $\Omega_0$ in 
the $(i+1)$-th iteration, $h_{i+1}$ and $(\Omega_0)_{i+1}$ are changed to 
\begin{eqnarray}
 h_{i+1} &=& h_i \delta + h'(1-\delta) \ , \\
 (\Omega_0)_{i+1} &=& (\Omega_0)_i\delta +\Omega'_0(1-\delta) \ ,
\end{eqnarray}
where $h'$ and $\Omega'_0$ are the quantities determined by  
Eqs.~(\ref{Eulereq2}) and~(\ref{jomega0}).  
The parameter 
$\delta$ ($0\leq \delta <1$) is used to control the changes of $h$ 
and $\Omega_0$ in an iteration 
step. For a very flattened configuration, we need to use $\delta >0.9$ 
to ensure convergence, and it takes more than 100 iterations for the 
models to converge to a fractional accuracy of $10^{-5}$. In the standard 
self-consistent field method, one only needs 
to solve for the density distribution $\rho$ (or equivalently, 
the enthalpy distribution $h$) self-consistently. In our self-consistent 
field method, we also need to solve for the equatorial angular velocity 
distribution $\Omega_0$ self-consistently. This 
is the main difference between the standard scheme and our proposed scheme, 
apart from the fact that the equations of motion in the 1PN case are 
more complicated.

The self-consistent field method described above computes stars with 
a given central energy density $\epsilon_c$ and equatorial radius 
$R_e$. However, we want to construct a star with a given 
total baryon mass $M_0$ and total angular momentum $J$. To do this, 
we first compute a model of non-rotating spherical star by solving 
the 1PN structure equations for nonrotating stars in isotropic coordinates. 
We use the density 
distribution as an initial guess to construct a model with slightly 
different $\epsilon_c$ and $R_e$. We then build models with different 
values of $\epsilon_c$ and $R_e$ until we end up with the model having 
the desired baryon mass and angular momentum.

For a rapidly rotating configuration, the equatorial radius extends 
to $R_e>1000~\rm{km}$ and the polar radius is approximately $R_p \approx 
10~\rm{km}$ in our coordinate system. Hence we use a nonuniform 
cylindrical grid to perform 
most of the computations. The resolution near the center of the star 
is about $0.4~\rm{km}$, whereas the resolution is about $6.5~\rm{km}$ 
near the equatorial surface of the star. We double the resolution to 
check the convergence. For a given $\epsilon_c$ and $R_e$, the fractional 
differences of the baryon mass $M_0$ and angular momentum $J$ between 
the two resolution grids are less than $10^{-5}$ even for the rapidly 
rotating cases. 

We adopt the Bethe-Johnson EOS~\cite{bethe74} for densities above
$10^{14}~\rm{g}~\rm{cm}^{-3}$, and BBP EOS~\cite{baym71} for densities
in the range $10^{11}-10^{14}~\rm{g}~\rm{cm}^{-3}$. The EOS for densities
below $10^{11}~\rm{g}~\rm{cm}^{-3}$ is joined by that of the pre-collapse
white dwarfs, which is the EOS of a zero-temperature ideal degenerate 
electron gas with electrostatic corrections derived by 
Salpeter~\cite{salpeter61}. We are mainly interested in the structure of 
the most rapidly rotating neutron stars. The central densities of these stars are 
around $4\times 10^{14}~\rm{g}~\rm{cm^{-3}}$ (see the next section), and 
ideas about the EOS in this relatively low density region have not 
changed very much since 1970's. 

The baryon masses of the neutron stars we compute 
in this chapter are around $1.4M_{\odot}$. For a non-rotating spherical 
star of this baryon mass, $c^2R/GM \approx 6$ for the EOS we adopt. Here 
$M\approx 1.3M_{\odot}$ is the gravitational mass and $R\approx 12~\rm{km}$ 
is the circumferential radius of the star. Hence we expect that the 
second and higher order post-Newtonian terms will give about $1/6^2\approx$3\% 
corrections to our models.

The effect of higher post-Newtonian terms can also be estimated by 
the virial identity to 1PN order (see the Appendix). We define a 
dimensionless quantity 
\begin{equation}
  \zeta=\frac{Z}{W_N} \ , 
\label{eq:Vtest}
\end{equation}
where 
\begin{eqnarray}
  Z &=& \int [ \sigma v^2+3P-\sigma (\varpi \partial_{\varpi} 
U + z\partial_z U)]\, d^3x \cr
   & & + \frac{1}{c^2} \int [6PU+\rho \Omega_0 \varpi Q+ 
(2P-\sigma v^2-4\rho U)(\varpi \partial_{\varpi} U+z\partial_z U) \cr
  & & + \rho\varpi \Omega_0 (\varpi \partial_{\varpi} Q+z\partial_z Q)]\, 
d^3 x \ , \\
  W_N &=&-\int \sigma (\varpi \partial_{\varpi}U + z\partial_z U)\, 
d^3 x \ , \\
  \sigma &=& \rho \left[ 1+\frac{1}{c^2}\left(v^2-2U+\Pi+\frac{P}{\rho}
\right)\right] \ .
\end{eqnarray}
The dimensionless quantity $\zeta$ is a measure of the fractional 
error of our equilibrium models due to higher post-Newtonian 
corrections (see the Appendix).

\section{\label{sec:results}Results}

We only construct neutron star models corresponding to the collapse 
of O-Ne-Mg white dwarfs (i.e., the Sequence III white dwarfs in 
Paper~II), because these neutron stars are the most likely
to undergo a dynamical instability and emit strong gravitational waves. 

\begin{figure}
\includegraphics[width=7cm]{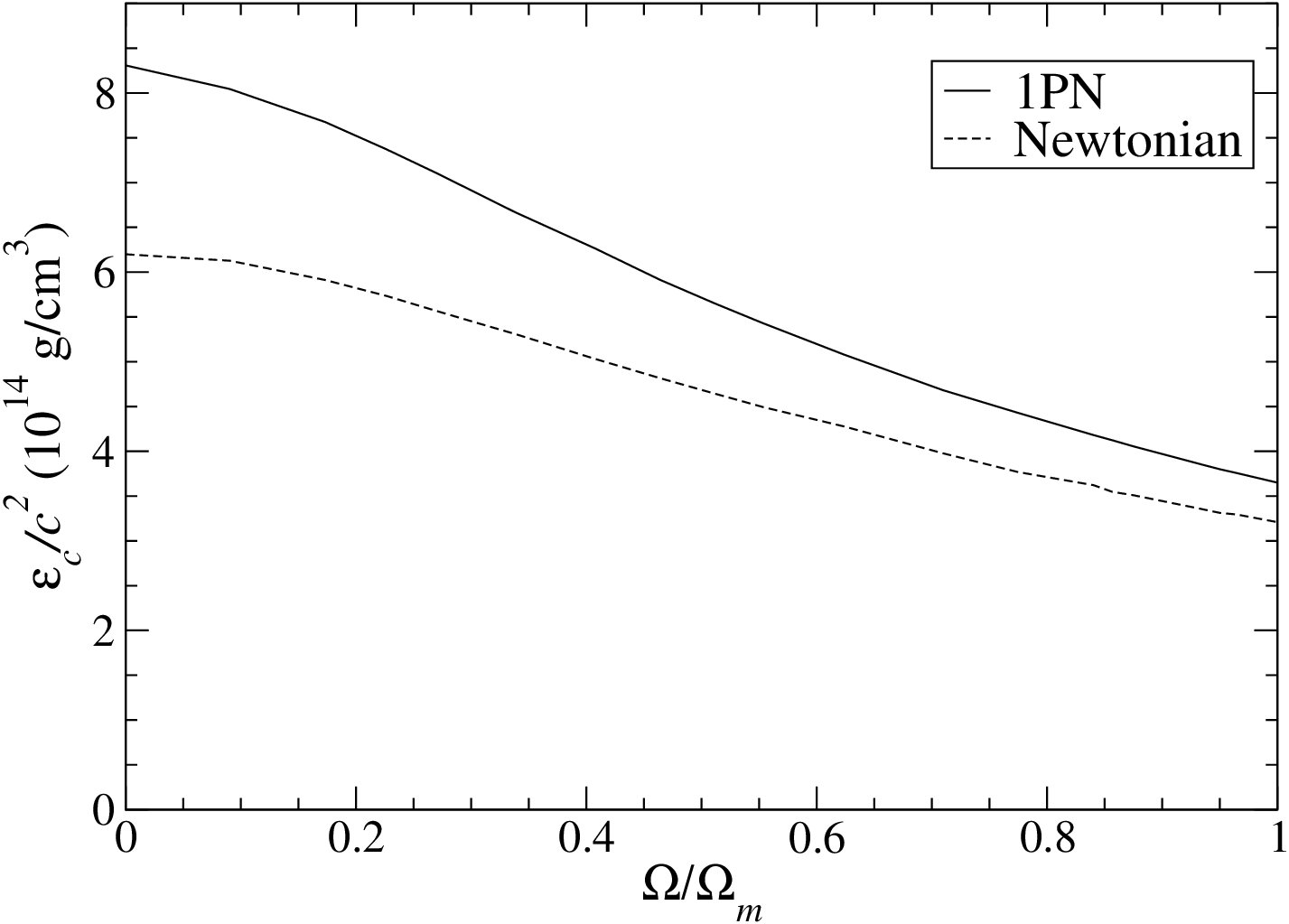}
\caption{\label{fig:ec}The central densities $\epsilon_c/c^2$ of 
differentially rotating 
neutron stars as a function of $\Omega/\Omega_m$ of the pre-collapse 
white dwarfs. Both Newtonian and 1PN results are shown for stars 
having the same $M_0$ and $J$.}
\end{figure}

Figure~\ref{fig:ec} shows the central densities $\epsilon_c/c^2$ of 
the resulting neutron stars as a function of $\Omega/\Omega_m$, 
where $\Omega$ is the angular frequency of the pre-collapse white 
dwarf, and $\Omega_m$ is the angular frequency of the maximally rotating 
white dwarf in the sequence. Both Newtonian and 1PN results are shown 
for stars having the same $M_0$ and $J$. 
We see that the central energy densities for the 1PN models are 
higher than the Newtonian models. This is expected because relativistic 
effects tend to make the stars more compact. The difference in $\epsilon_c$ 
decreases as the star becomes more rapidly rotating. 

\begin{figure}
\includegraphics[width=7cm]{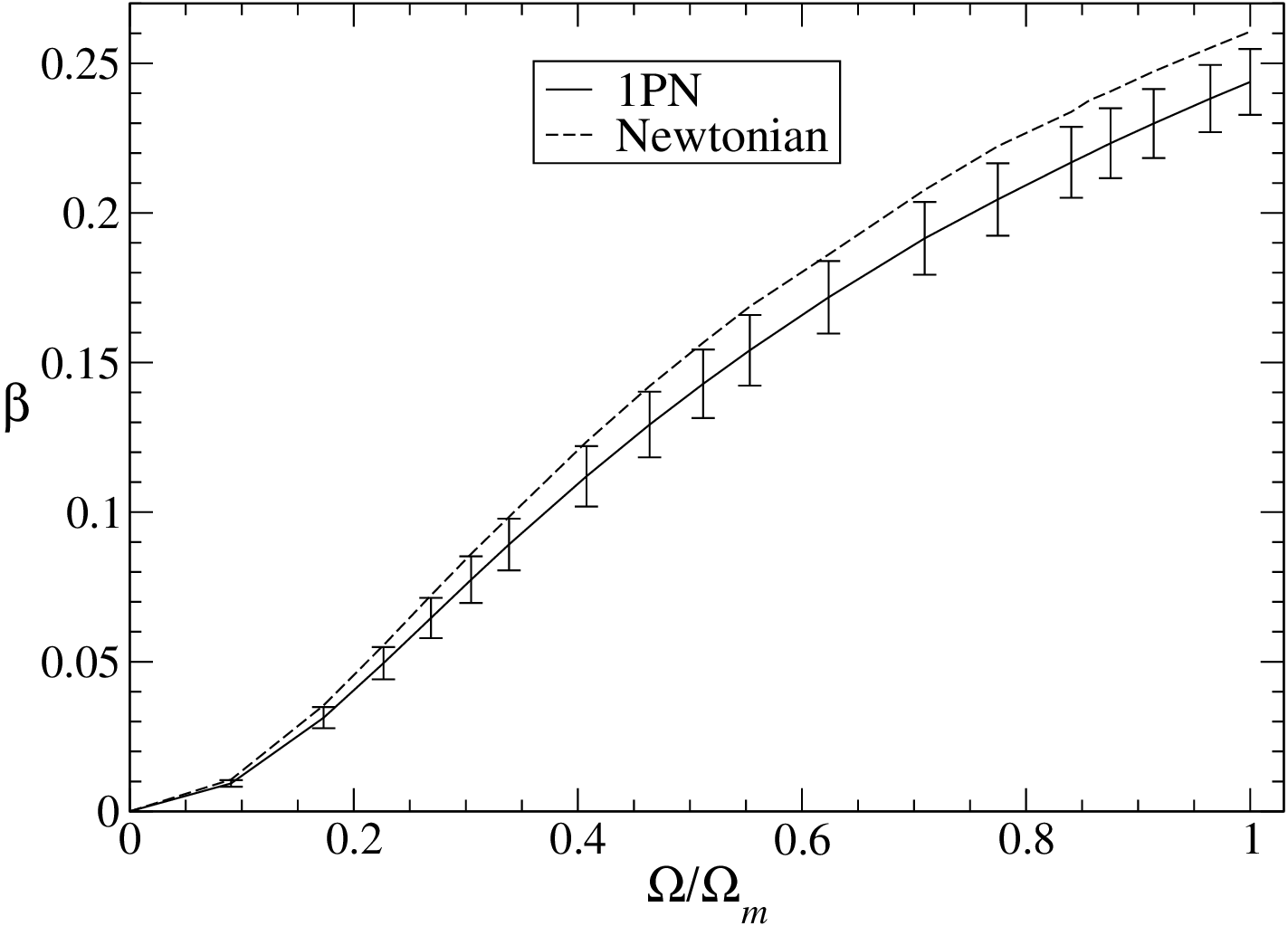}
\caption{\label{fig:beta}The value of $\beta$ of the resulting neutron 
stars as a function of $\Omega/\Omega_m$ of the pre-collapse white dwarfs. 
The vertical bars through the 1PN curve show $\beta\pm \delta \beta$ of 
selected equilibrium models, where $\delta \beta$ is the error 
of $\beta$ due to higher post-Newtonian corrections estimated by 
Eq.~(\ref{eq:deltabeta}).}
\end{figure}

Figure~\ref{fig:beta} shows the value of $\beta=T/|W|$ of the neutron 
stars as a function of $\Omega/\Omega_m$ for both the Newtonian 
and 1PN models. To estimate the possible error 
of $\beta$ due to higher post-Newtonian effects, we use the formula 
\begin{eqnarray}
  \delta \beta &=& \beta \left| \frac{\delta T}{T}-\frac{\delta W}{W}\right| \cr
  &\leq & \beta \left( \left| \frac{\delta T}{T}\right| + 
\left|\frac{\delta W}{W}\right| \right) \ . \label{eq:deltabeta}
\end{eqnarray}
We found that the quantity $U^2/c^4$ is the largest second post-Newtonian 
terms we neglected in the whole calculation. Hence we estimate that 
\begin{eqnarray}
  \left| \frac{\delta T}{T}\right| &\approx & \frac{1}{T} \int T_i \frac{U^2}{c^4}\, 
d^3x \ , \\
  \left| \frac{\delta W}{W}\right| &\approx & \left| \frac{1}{W} 
\int W_i \frac{U^2}{c^4}\, d^3x \right|  \ , \label{eq:dlogW}
\end{eqnarray}
where $T_i$ and $W_i$ are the integrands in Eqs.~(\ref{eq:pnT}) 
and~(\ref{eq:pnW}), respectively. The vertical bars in Fig.~\ref{fig:beta} 
show $\beta\pm \delta \beta$ of selected equilibrium models [using 
Eqs.~(\ref{eq:deltabeta})--(\ref{eq:dlogW}) for $\delta \beta$]. The 
result suggests that relativistic effect lowers the value of $\beta$ for 
stars of given $M_0$ and $J$. 
The maximum $\beta$ of these neutron stars is 0.24, which is 8\% lower 
than the Newtonian case (0.26). However, the error bars also suggest that 
higher post-Newtonian corrections could change the values of $\beta$ 
significantly. 

\begin{figure}
\includegraphics[width=7cm]{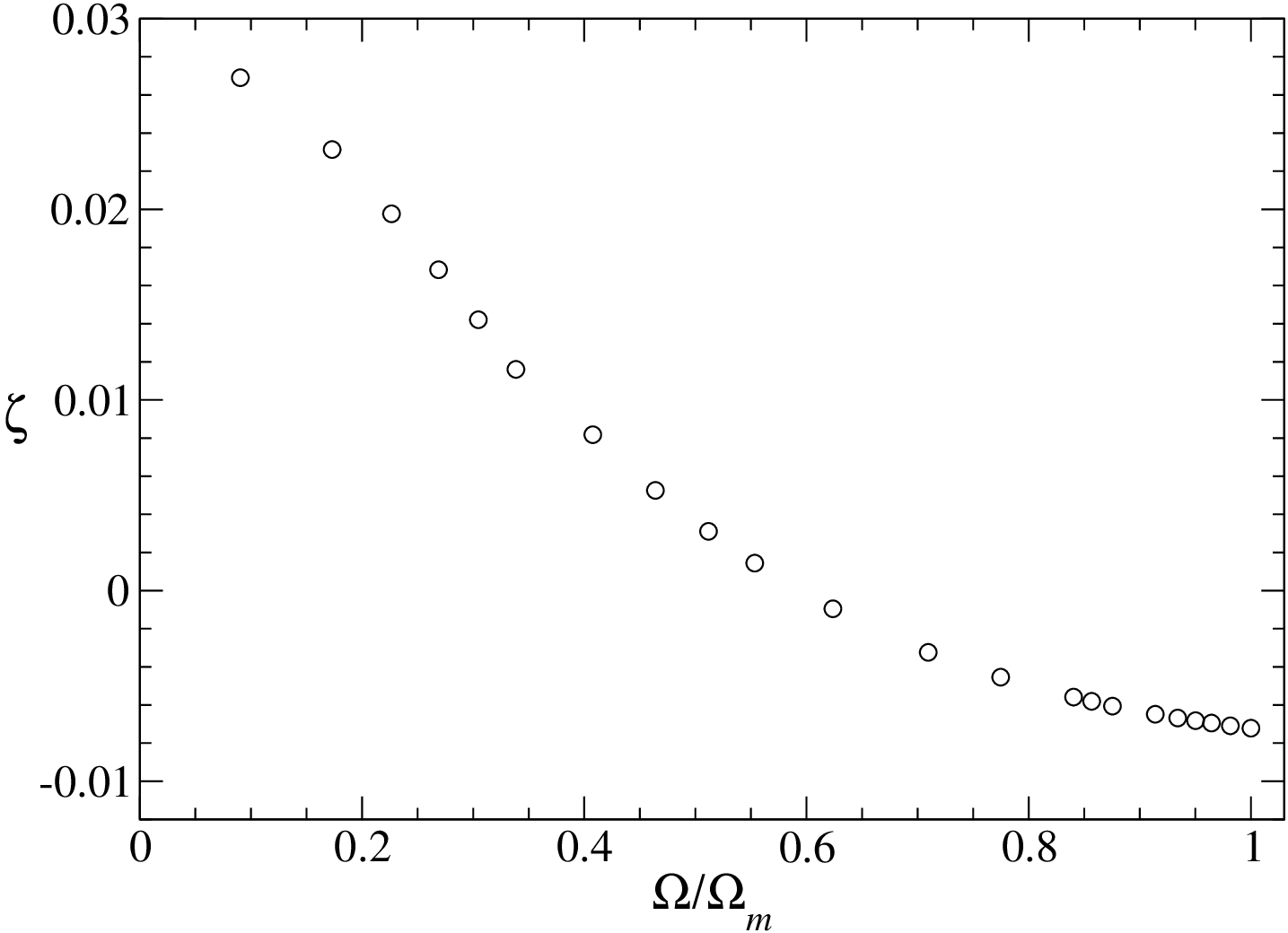}
\caption{\label{fig:Vtest}The virial quantity $\zeta$ defined in 
Eq.~(\ref{eq:Vtest}) for the equilibrium neutron star models, parametrized 
by $\Omega/\Omega_m$ of the pre-collapse white dwarfs.}
\end{figure}

Figure~\ref{fig:Vtest} shows the virial quantity $\zeta$ for our 
equilibrium neutron star models. This quantity is a measure of the fractional 
correction to the equilibrium structure of the stars due to higher post-Newtonian 
effects. We see that $\zeta$ is smaller than 0.03 for all the models we computed, 
which is in accord with our estimate near the end of Sec.~\ref{sec:num}.

\begin{figure}
\includegraphics[width=7cm]{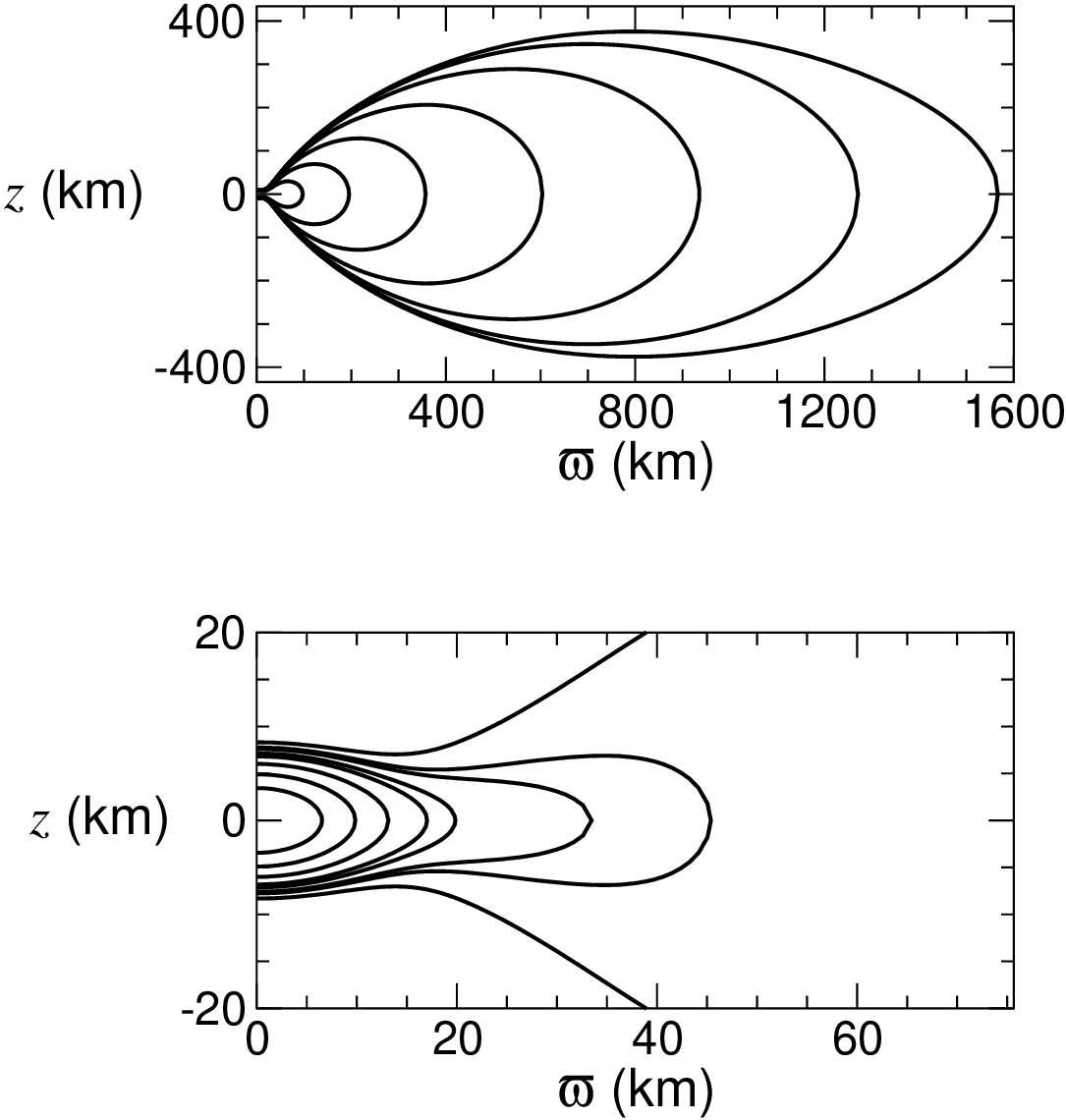}
\caption{\label{fig:dencont}Meridional density contours of the neutron 
star resulting from 
the AIC of a rigidly rotating O-Ne-Mg white dwarf with $\Omega/\Omega_m=0.964$. 
This neutron star has $\beta=0.238$. The contours in the upper graph denote, 
from inward to outward, $\epsilon/\epsilon_c=
10^{-4}$, $10^{-5}$, $10^{-6}$, $10^{-7}$, $10^{-8}$, $10^{-9}$ and 0.
The contours in the lower graph denote, from inward to outward, $\epsilon/\epsilon_c$=
0.8, 0.6, 0.4, 0.2, 0.1, $10^{-2}$, $10^{-3}$ and $10^{-4}$. The
central density of the star is $\epsilon_c/c^2=3.8\times 10^{14}~\rm{g}~\rm{cm}^{-3}$.}
\end{figure}

The structure of the neutron stars is not much different from the 
Newtonian models. Stars with $\beta \agt 0.1$ all contain a high-density 
central core of size about 20~km, surrounded by a 
low-density torus-like envelope. The size of the envelope ranges from 
100~km (for stars with $\beta \sim 0.1$) to over 500~km (for $\beta \agt 0.2$). 
Figure~\ref{fig:dencont} shows the density contour of a typical rapidly 
rotating neutron star. This figure looks basically the same as 
Fig.~3 of Paper~II, which shows the density contours of the same star 
computed with Newtonian gravity. The $\beta$ of this star is 0.238, which is 
somewhat smaller than the Newtonian value 0.255. 

\begin{figure}
\vskip 1cm
\includegraphics[width=7cm]{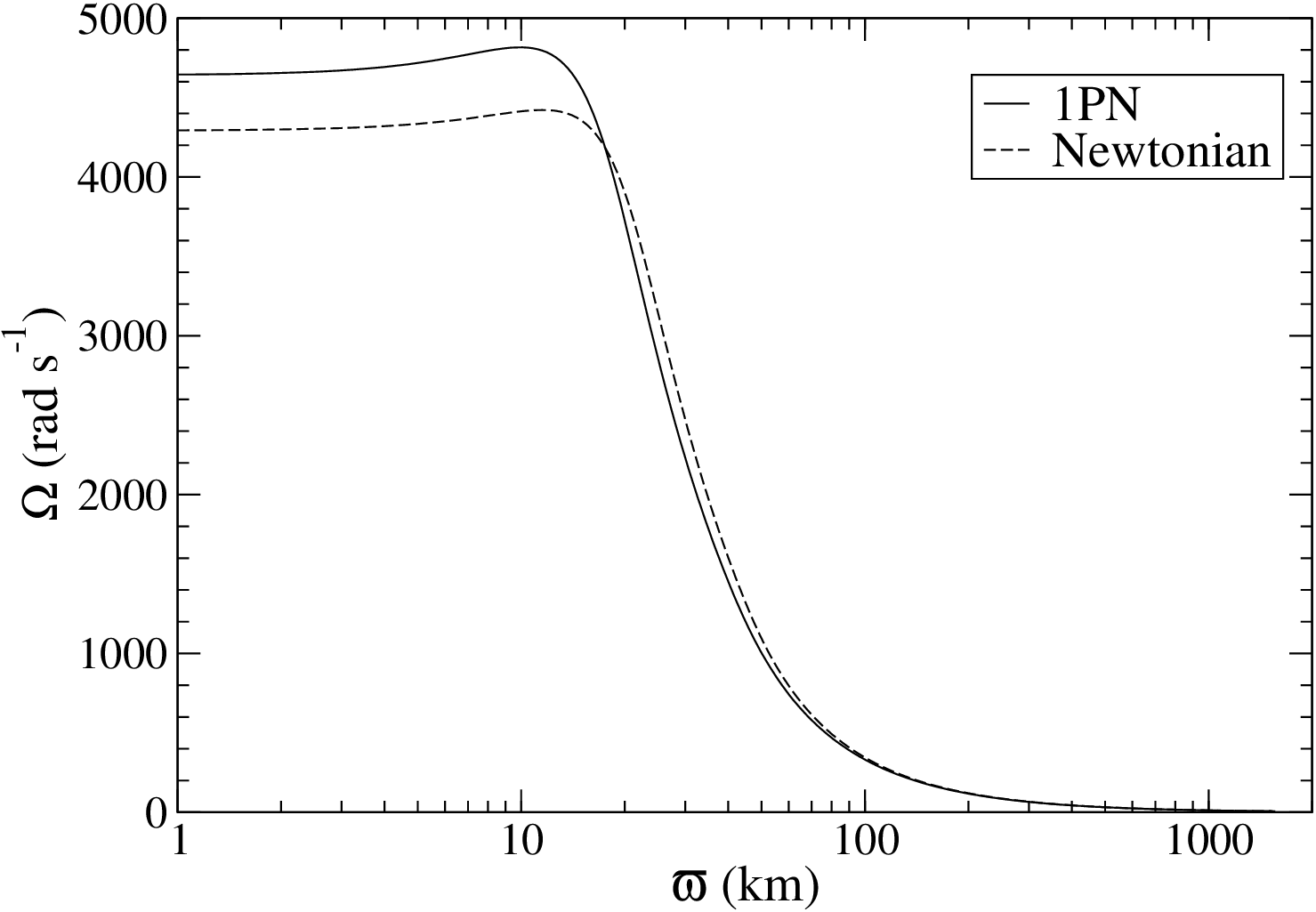}
\caption{\label{fig:om2156}The equatorial angular velocity $\Omega_0(\varpi)$ of the
neutron star in Fig.~\ref{fig:dencont}.}
\end{figure}

\begin{figure}
\vskip 1cm
\includegraphics[width=7cm]{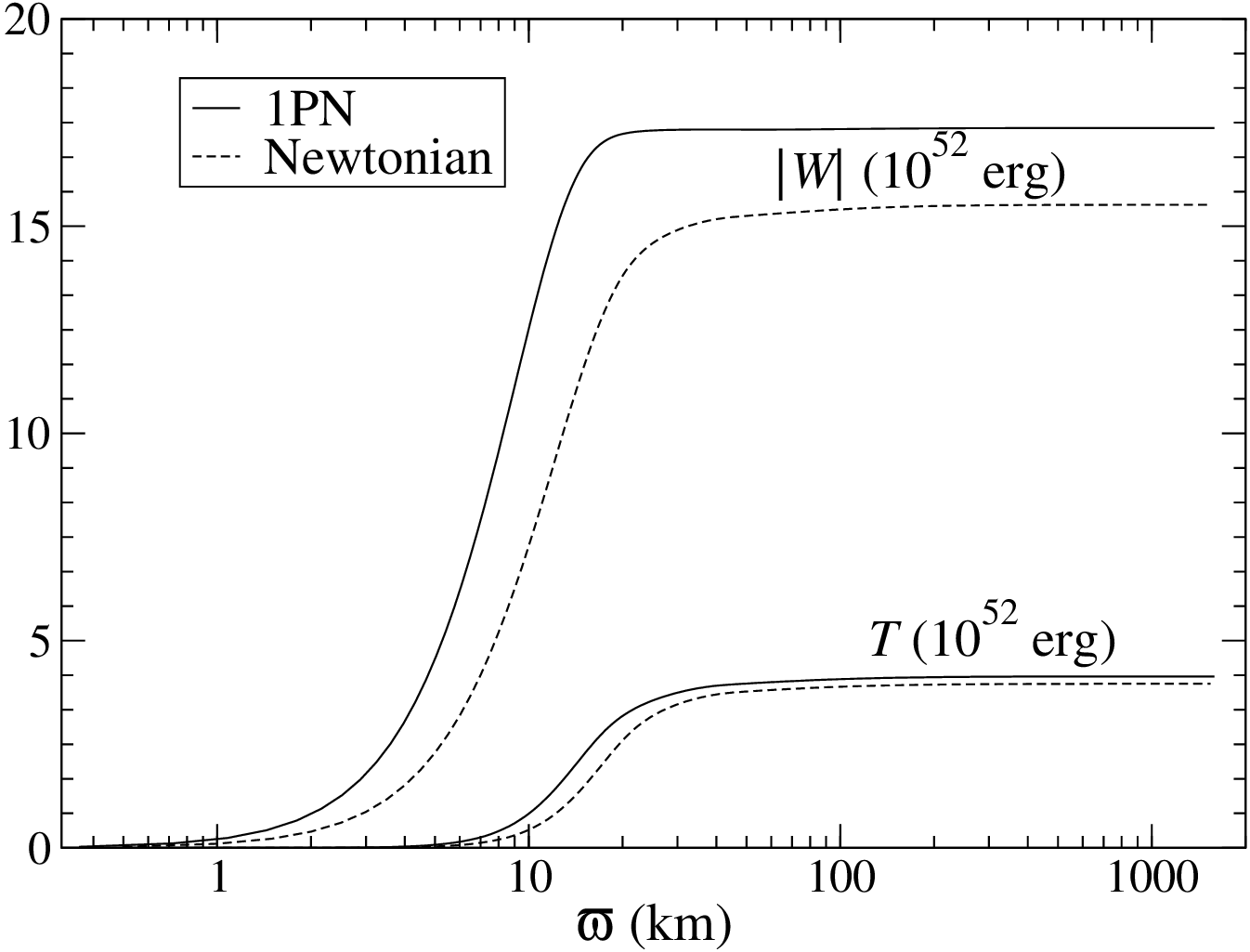}
\caption{\label{fig:energies}The distribution of the rotational kinetic 
energy $T$ and gravitational 
binding energy $|W|$ of the material inside the radius $\varpi$, for the neutron
star in Fig.~\ref{fig:dencont}.}
\end{figure}

Figure~\ref{fig:om2156} shows 
the equatorial angular velocity distribution $\Omega_0(\varpi)$ of the 
same star. We see that the angular velocity in the inner core of the star 
($\varpi \alt 20~\rm{km}$) in the 1PN model is slightly larger than that of 
the corresponding Newtonian model. This is expected 
because relativistic effects make the star more compact. The material 
is compressed more in the 1PN model, and should rotate faster 
due to the conservation of angular momentum. Figure~\ref{fig:energies} 
shows the distribution of rotational kinetic energy $T$ and gravitational 
binding energy $|W|$ of the material contained within cylindrical radius 
$\varpi$. The two 
quantities approach their asymptotic values at $\varpi \approx 30~\rm{km}$. 
This is due to the high central condensation of the star. Both $T$ and 
$|W|$ in the 1PN model are larger than the corresponding Newtonian 
model. The kinetic energy $T$ is larger because the star rotates 
faster. However, the difference between the two $T$-curves decreases 
as we move away from the rotation axis. This is because most of the 
kinetic energy of the star is from the region 
$10~\rm{km}\alt \varpi \alt 30~\rm{km}$, in which relativistic effects 
are less important. On the other hand, the gravitational binding energy 
is mainly contributed from the material in the inner region 
$\varpi \alt 20~\rm{km}$, in which 
relativistic effects are important. As a result, the $T/|W|$ value 
of the relativistic model is somewhat less than in the corresponding Newtonian 
model.

\begin{figure}
\vskip 1cm
\includegraphics[width=7cm]{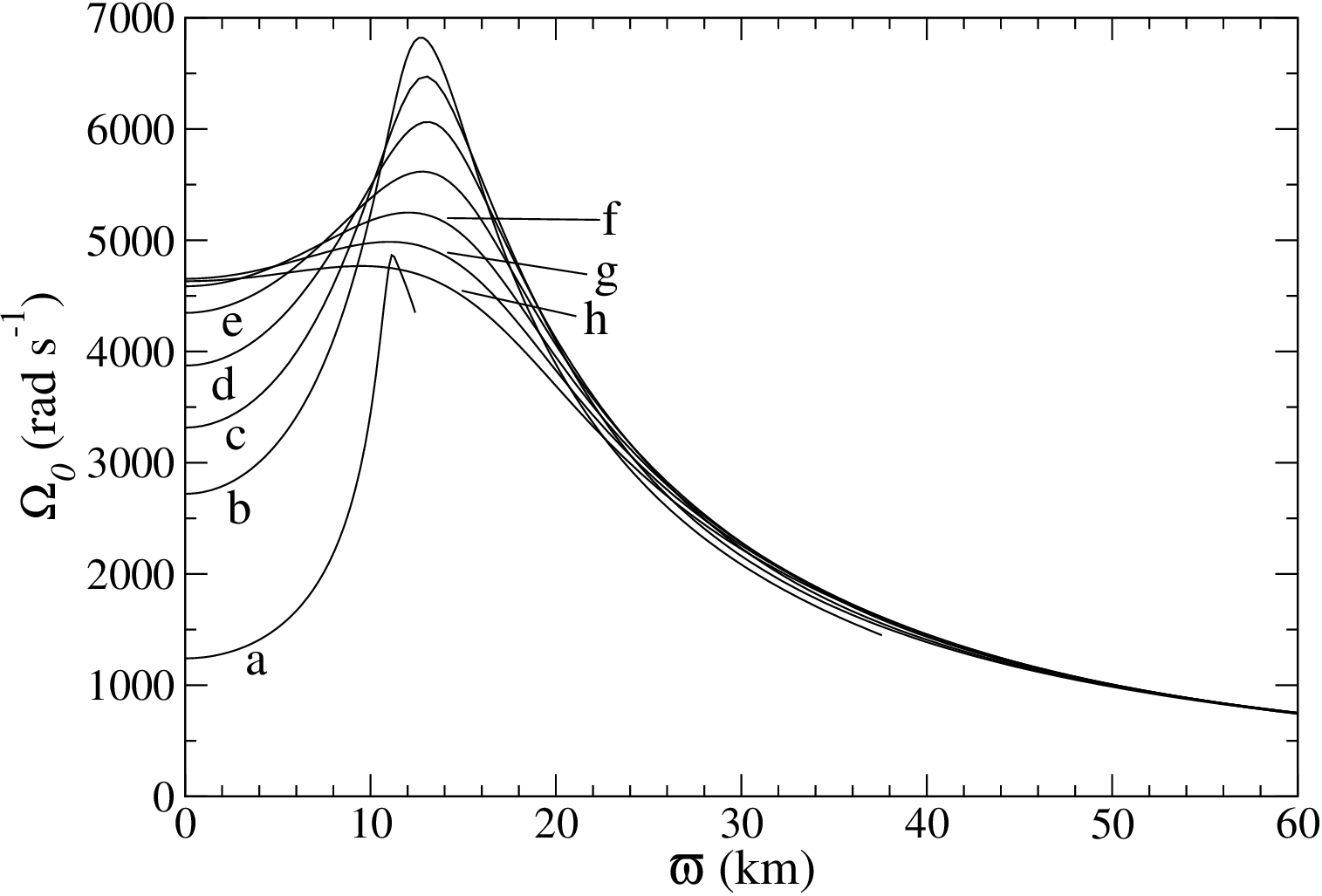}
\caption{\label{fig:omega} The equatorial rotational angular velocity $\Omega_0$ 
as a function of $\varpi$ for $\varpi<60~\rm{km}$. These neutron stars 
result from the AIC of the pre-collapse white dwarfs with $\Omega/\Omega_m$ 
equal to (a)~0.090, (b)~0.23, (c)~0.30, (d)~0.41, (e)~0.55, (f)~0.71, 
(g)~0.86 and (h)~1.00.}
\end{figure}

\begin{figure}
\vskip 1cm
\includegraphics[width=7cm]{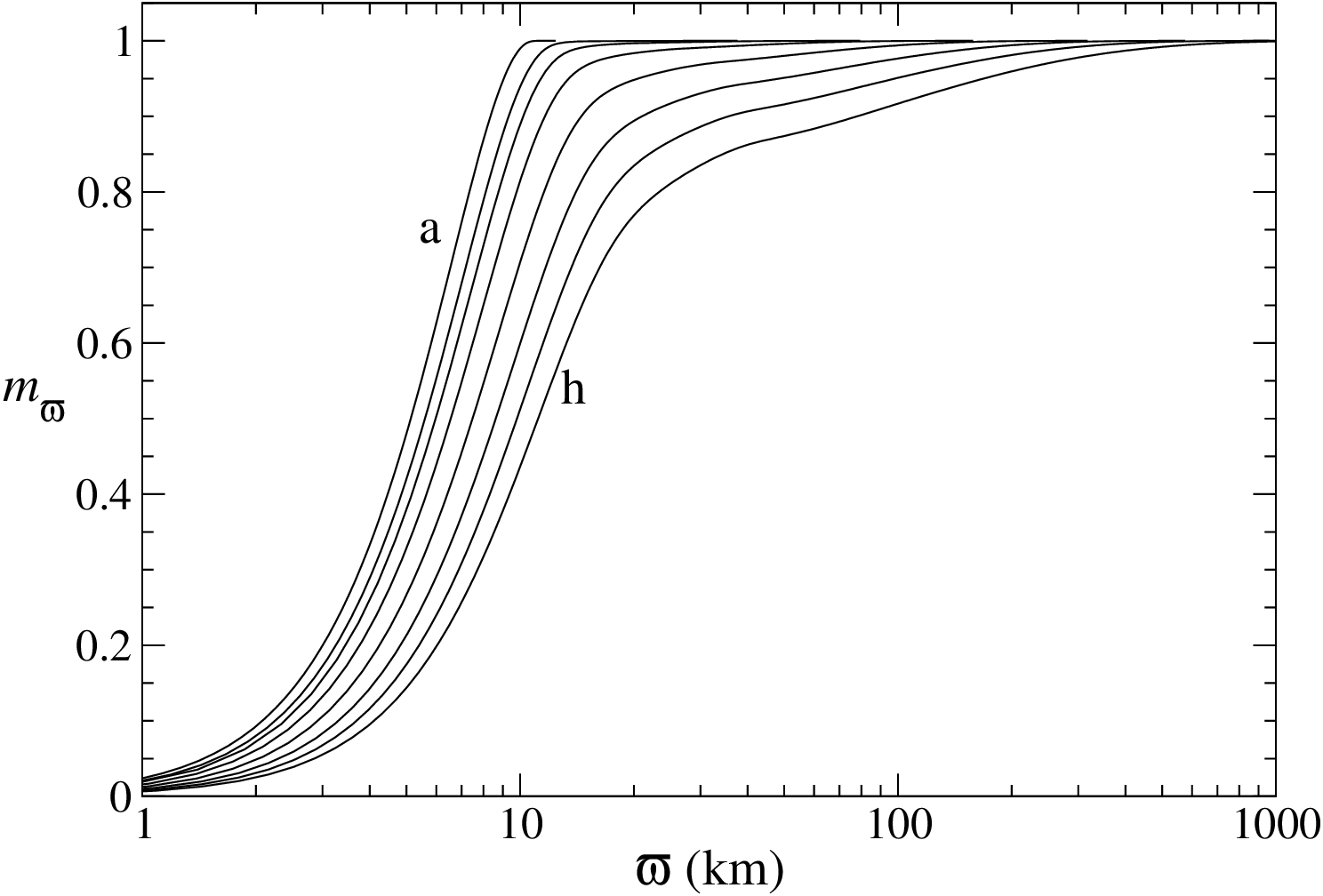}
\caption{\label{fig:mp}The cylindrical mass 
fraction $m_{\varpi}$ as a function of 
$\varpi$ for the neutron star models in~Fig.~\ref{fig:omega}. The 
curves and the corresponding models are identified by the degrees of 
central condensation: the higher the degree of central condensation, 
the lower the value of $\Omega/\Omega_m$.}
\end{figure}

\begin{figure}
\vskip 1cm
\includegraphics[width=7cm]{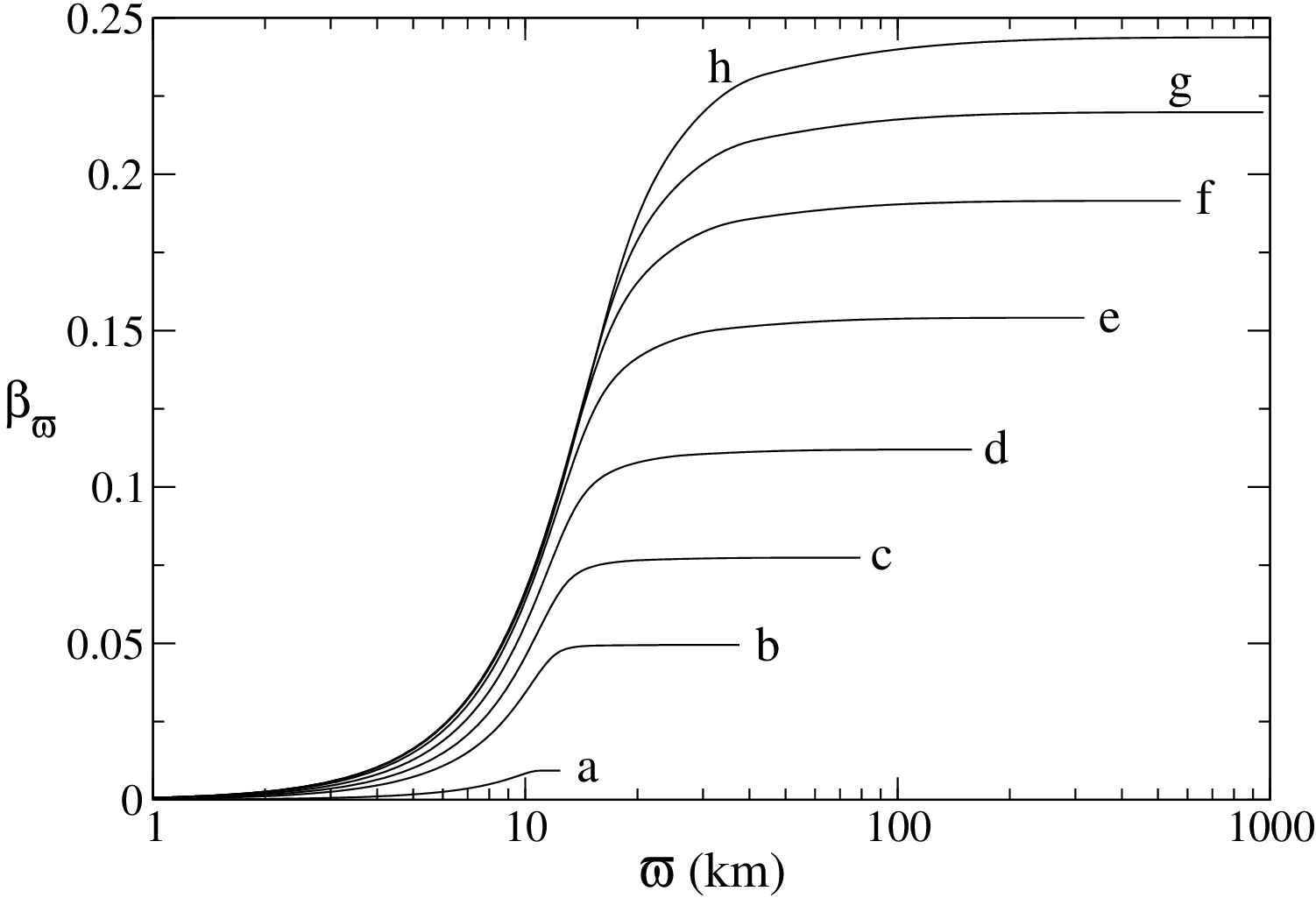}
\caption{\label{fig:betapm}The value of $\beta_{\varpi}$ as a function 
of $\varpi$ for the neutron star models in Fig.~\ref{fig:omega}.} 
\end{figure}

Figure~\ref{fig:omega} shows the equatorial angular velocity $\Omega_0$ 
for several selected models in the central region near the rotation axis. The 
shape of the curves are very similar to those of the Newtonian models
(see Fig.~4 of Paper~II). However, the angular velocities in the 1PN 
models are all slightly larger than the Newtonian models in the inner core. 

Figure~\ref{fig:mp} shows the baryonic mass fraction $m_{\varpi}$ 
verses $\varpi$ for the selected models in Fig.~\ref{fig:omega}. 
As in the Newtonian case (see Fig.~5 of Paper~II), 
the mass is highly concentrated in the inner core of the star. The 
degree of central condensation decreases as the star rotates faster. However, 
more than 80\% of the mass is contained within a 30~km radius 
even for the most rapidly rotating star, where the outer envelope 
extends to over 1000~km. The collapsed object can be regarded as a neutron star 
of size about 20~km surrounded by an accretion torus. 

Figure~\ref{fig:betapm} shows $\beta_{\varpi}$, the $T/|W|$ of the material 
inside the surface of constant $\varpi$, for the selected models in 
Fig.~\ref{fig:omega}. The shape of the curves are qualitatively the 
same as those in the Newtonian models ( see Fig.~6 of Paper~II), 
although the values of $\beta_{\varpi}$ are slightly smaller. All the 
curves level off at $\varpi \approx 30 - 50~\rm{km}$, suggesting that 
the material in the outer layers does not have much influence on the overall 
dynamical stability of the star.

\section{\label{sec:con}Conclusions}

We have generalized the self-consistent field method so that it can be used 
to compute models of differentially rotating stars to 1PN order with 
a specified angular momentum distribution $j(m_{\varpi})$. We also 
applied this new method to construct models of nascent neutron stars 
resulting from the collapse of massive O-Ne-Mg white dwarfs we studied 
in Paper~II and compare them with the corresponding Newtonian models.

We found that the 1PN models are more compact and rotate faster. Our 
calculations also suggest that the 1PN models have 
smaller values of $\beta$ than the corresponding Newtonian models. 
The highest value of $\beta$ these neutron stars can achieve is 0.24, which 
is 8\% smaller than the Newtonian case. Higher post-Newtonian 
corrections may change the values of $\beta$, but our estimate 
suggests that they will still be smaller than those of the Newtonian models. 
We estimate that the fractional error of our 1PN models 
due to our neglecting higher order post-Newtonian terms is about 3\%. 

The fact that relativistic effects lower the values of $\beta$ can be 
understood by the following heuristic argument. Relativistic effects 
make a star more compact. Hence the gravitational binding enetgy $|W|$ 
increases. The compactness of the star also make it 
rotate faster due to conservation of angular momentum. Hence the 
rotational kinetic energy $T$ also increases. However, most of the 
kinetic energy comes from material slightly away from the center 
of the star (in the region $10~{\rm km}\alt \varpi \alt 30~{\rm km}$ for 
the star in Fig.~\ref{fig:energies}), where relativistic effects are less 
important compared to the region near the center. On the other hand, 
the gravitational binding energy $|W|$ is contributed mainly from 
material near the center of the star (in the region $\varpi \alt 20~{\rm km}$ 
for the star in Fig.~\ref{fig:energies}). As a result, the increase in $T$ is not 
as much as the increase in $|W|$, and so the value of $\beta$ decreases.

We have demonstrated that relativistic effects could lower the value of 
$\beta$ of a star with a given baryon mass $M_0$ and angular momentum $J$. 
Shibata, Baumgarte, Shapiro and Saijo~\cite{shibata00,saijo00} demonstrated 
that relativistic effects also lower the critical value $\beta_d$ 
for the dynamical instability by a similar amount. It will be interesting 
to find out which of these two effects is more important. Careful 
numerical 1PN stability analyses must be carried out to determine whether 
or not relativistic effects destabilize the stars.

\begin{acknowledgments}
I thank Lee Lindblom and Kip Thorne for useful discussions and comments on
various aspects of this work. I also thank Thomas Baumgarte, Matthew Duez,
Pedro Marronetti, Stuart Shapiro for useful discussions. This research 
was supported by NSF grants PHY-9796079 and PHY-0099568.
\end{acknowledgments}

\appendix*

\section{Virial theorem to 1PN order}

Virial theorems in full general relativity were formulated by 
Bonazzola and Gourgoulhon~\cite{bonazzola73,gourgoulhon94,bonazzola94}, and 
have been proved to be useful as a consistency check for numerical computation 
of rotating star models (see, e.g.,~\cite{bonazzola74,bonazzola93,cook96}). 
These virial identities are not very convenient to implement since they 
involve integrals over the entire two or three-dimensional volume, but 
our computational domain only extends to the region slightly outside the 
surface of the star. Chandrasekhar derived a virial identify 
to the 1PN order~\cite{chandra65} in which integrations are 
only performed over the interior of a star. However, it involves a double volume 
integral. In this appendix, we shall derive another virial identify which is 
correct to the 1PN order and is easier to compute than the Chandrasekhar 
identity. This expression is useful to estimate the error of our equilibrium 
neutron star models due to higher post-Newtonian effects.

We start with Eq.~(67) of Ref.~\cite{chandra65}, which is the 
Euler equation to 1PN order~\cite{note}: 
\begin{eqnarray}
  & \partial_i (\sigma v_j v^i) + \left(1+\frac{2U}{c^2}\right) \partial_j P +
\sigma \left(1+\frac{v^2}{c^2}\right) \partial_j U 
  -\frac{4}{c^2}\rho v_j v^i \partial_i U & \cr 
& +\frac{1}{c^2}\rho v^i(\partial_i A_j- \partial_j A_i) +\frac{4\rho U}{c^2}
\partial_j U =0 & \ , \label{eq:Euler1pnch} \\
  & \sigma = \rho \left[ 1+ \frac{1}{c^2}\left(v^2-2U+\Pi+\frac{P}{\rho}\right)
\right] & \ .
\end{eqnarray}
Here we have set all partial time derivatives to zero. The equation is 
valid to 1PN order for the metric~(\ref{g00})--(\ref{gij}). Multiplying 
Eq.~(\ref{eq:Euler1pnch}) by $x^j$, summing over $j$ and integrating over 
$d^3x \equiv dx\, dy\, dz=\varpi\, d\varpi \, dz \, d\varphi$, we obtain, 
after integration by parts, 
\begin{eqnarray}
   \int \{ \sigma v^2+3P-\sigma x^j \partial_j U
+\frac{1}{c^2}[ 2P(3U+x^j\partial_j U) - \sigma v^2 x^j\partial_j U + 
4\rho x^j v_j v^i \partial_i U & & \cr
  -\rho v^i x^j (\partial_i A_j - \partial_j A_i)-4\rho U x^j \partial_j U] \} \,
d^3x = 0 \label{eq:virialpn} & & \ .
\end{eqnarray}
In a stationary, axisymmetric and circular spacetime, $x^j v_j =  x^j A_j=0$. 
In cylindrical coordinates, we have 
\begin{eqnarray}
  x^j \partial_j U &=& \varpi \partial_{\varpi} U + z \partial_z U \ , \\
  v^i x^j (\partial_j A_i-\partial_i A_j) &=& \varpi \Omega (\varpi \partial_{\varpi} 
Q + z \partial_z Q)+\varpi \Omega Q \ , 
\end{eqnarray}
where $Q=(xA_y-yA_x)/\varpi$. Hence Eq.~(\ref{eq:virialpn}) becomes, to 1PN order, 
\begin{eqnarray}
  \int \{ \sigma v^2+3P-\sigma (\varpi \partial_{\varpi}
U + z\partial_z U) + \frac{1}{c^2} [6PU+\rho \Omega_0 \varpi Q \hspace{5cm} & & \cr
+(2P-\sigma v^2-4\rho U)(\varpi \partial_{\varpi} U+z\partial_z U) 
+ \rho\varpi \Omega_0 (\varpi \partial_{\varpi} Q+z\partial_z Q)] \} \, 
d^3 x = 0 \ . \hspace{1cm} & & \label{eq:virialpn2}
\end{eqnarray}
Equation~(\ref{eq:virialpn2}) is our virial identity. It involves 
a volume integral over the interior of the star. In the Newtonian limit, 
it reduces to the familiar form $2T+3\Lambda+W=0$, where $\Lambda=\int P\, d^3x$.

Let $Z$ be the left side of Eq.~(\ref{eq:virialpn2}) and define a dimensionless 
quantity 
\begin{eqnarray}
  \zeta &=& Z/W_N \ \ \ , \\
  W_N &=&-\int \sigma (\varpi \partial_{\varpi}U + z\partial_z U)\, d^3 x \ .
\end{eqnarray}
In the Newtonian limit, $\zeta=(2T+3\Lambda+W)/W$, which is a quantity often 
used as a measure of the fractional numerical error of Newtonian equilibrium 
models caused by finite grid size. 

Our equilibrium models are constructed by solving Eq.~(\ref{Eulereq2}), which 
agrees with Eq.~(\ref{eq:Euler1pnch}) to 1PN order. Hence the quantity $\zeta$ 
measures the fractional error of our models due to second 
and higher post-Newtonian corrections, and also the numerical error due to 
finite grid size.


\begin{thebibliography}{99}

\bibitem{liu02} Y.T.\ Liu, Phys.\ Rev.\ D., \textbf{65}, 124003 (2002)
(Paper~II).

\bibitem{liu01} Y.T. Liu and L.\ Lindblom, Mon.\ Not.\ R.\
Astro.\ Soc., \textbf{324}, 1063 (2001) (Paper~I).

\bibitem{shibata00} M.\ Shibata, T.W.\ Baumgarte and S.L.\ Shapiro,
Astrophys.\ J., \textbf{542}, 453 (2000).

\bibitem{saijo00} M.\ Saijo, M.\ Shibata, T.W.\ Baumgarte and S.L.\ Shapiro,
Astrophys.\ J., \textbf{548}, 919 (2001).

\bibitem{wilson72} J.R.\ Wilson, Astrophys.\ J., \textbf{176}, 195 (1972). 

\bibitem{bonazzola74} S.\ Bonazzola and J.\ Schneider, Astrophys.\ J., 
\textbf{191}, 273 (1974).

\bibitem{butterworth75} E.M.\ Butterworth and J.R.\ Ipser, 
Astrophys.\ J.\ Lett., \textbf{200}, 103 (1975); 

\bibitem{butterworth76} E.M.\ Butterworth and J.R.\ Ipser,
Astrophys.\ J., \textbf{204}, 200 (1976).

\bibitem{butterworth7679} E.M.\ Butterworth, Astrophys.\ J., \textbf{204}, 
561 (1976); E.M.\ Butterworth, Astrophys.\ J., \textbf{231}, 219 (1979).

\bibitem{friedman86} J.L.\ Friedman, J.R.\ Ipser and L.\ Parker, 
Astrophys.\ J., \textbf{304}, 115 (1986). 

\bibitem{komatsu89} H.\ Komatsu, Y.\ Eriguchi and I.\ Hachisu, 
Mon.\ Not.\ R.\ Astro.\ Soc., \textbf{237}, 355 (1989); 
H.\ Komatsu, Y.\ Eriguchi and I.\ Hachisu, Mon.\ Not.\ R.\ Astro.\ Soc., 
\textbf{239}, 153 (1989). 

\bibitem{monchmeyer88} R.\ M\"{o}nchmeyer and E.\ M\"{u}ller, in NATO ASI
on \textit{Timing Neutron Stars}, ed.\ \"{O}gelman H., D.\ Reidel Publ.\
Comp., Dordrecht 1988.

\bibitem{janka89} H.-T.\ Janka, R.\ M\"{o}nchmeyer, Astro.\ \& 
Astrophys., \textbf{209}, L5 (1989);
H.-T.\ Janka, R.\ M\"{o}nchmeyer, Astro.\ \& Astrophys., \textbf{226}, 69 (1989).

\bibitem{fryer01} C.L.\ Fryer, D.E.\ Holz and S.A.\ Hughes, Astrophys.\ J., 
\textbf{565}, 430 (2002).

\bibitem{ostriker68} J.P.\ Ostriker and J. W-K.\ Mark, Astrophys.\ J., 
\textbf{151}, 1075 (1968);
J.P.\ Ostriker and P.\ Bodeneimer, Astrophys.\ J., \textbf{151}, 1089 (1968).

\bibitem{bodenheimer73} P.\ Bodeneimer and J.P.\ Ostriker, Astrophys.\ J., 
\textbf{180}, 159 (1973).

\bibitem{pickett96} B.K.\ Pickett, R.H.\ Durisen and G.A.\ Davis,
Astrophys.\ J., \textbf{458}, 714 (1996).

\bibitem{new01} K.C.B.\ New and S.L.\ Shapiro, Astrophys.\ J., 
\textbf{548}, 439 (2001).

\bibitem{bardeen70} J.M.\ Bardeen, Astrophys.\ J., \textbf{162}, 71 (1970).

\bibitem{chandra65} S.\ Chandrasekhar, Astrophys.\ J., \textbf{142}, 1488 (1965). 

\bibitem{blanchet90} L.\ Blanchet, T.\ Damour and G.\ Scha\"afer, 
Mon.\ Not.\ R.\ Astro.\ Soc., \textbf{242}, 289 (1990).

\bibitem{cutler91} C.\ Cutler, Astrophys.\ J., \textbf{374}, 248 (1991).

\bibitem{MTW} C.\ W.\ Misner, K.S.\ Thorne and J.A.\ Wheeler, 
\textit{Gravitation} (Freeman and Company 1973).

\bibitem{smith92} S.\ Smith, J.M.\ Centrella, in \textit{Approaches to 
Numerical Relativity}, ed.\ R.d'Inverno.\ (Cambridge Univ.\ Press, 
New York 1992).

\bibitem{hachisu86} I.\ Hachisu, Astrophys.\ J.\ Supp., \textbf{61}, 
479 (1986).

\bibitem{bethe74} H.A.\ Bethe and M.B.\ Johnson, Nucl.\ Phys.\ A, \textbf{230},
1 (1974).

\bibitem{baym71} G.\ Baym, H.A.\ Bethe and C.J.\ Pethick, Nucl.\ Phys. A,
\textbf{175}, 225 (1971).

\bibitem{salpeter61} E.E.\ Salpeter, Astrophys.\ J., \textbf{134}, 669 (1961).

\bibitem{bonazzola73} S.\ Bonazzola, Astrophys.\ J., \textbf{182}, 335 (1973). 

\bibitem{gourgoulhon93} E.\ Gourgoulhon and S.\ Bonazzola, Phys.\ Rev.\ 
D., \textbf{48}, 2635 (1993).

\bibitem{gourgoulhon94} E.\ Gourgoulhon and S.\ Bonazzola, Class.\ Quantum 
Grav., \textbf{11}, 443 (1994).

\bibitem{bonazzola94} S.\ Bonazzola and E.\ Gourgoulhon, Class.\ Quantum 
Grav., \textbf{11}, 1775 (1994).

\bibitem{bonazzola93} S.\ Bonazzola, E.\ Gourgoulhon, M.\ Salgado and 
J.A.\ Marck, Astron.\ Astrophys., \textbf{278}, 421 (1993).

\bibitem{cook96} G.B.\ Cook, S.L.\ Shapiro and S.A.\ Teukolsky, Phys.\ 
Rev.\ D., \textbf{53}, 5533 (1996).

\bibitem{note} The relationships between our variables and those used in 
Ref.~\cite{chandra65} are: $U=-(U^*+2\Phi^*/c^2)$ and $A_i=-4U_i^*$, 
where the variables appeared in Ref.~\cite{chandra65} are denoted by 
the superscript $^*$.

\end{thebibliography}
\end{document}